
\input phyzzx
\nopubblock

\bigskip

\def\mht{\tilde T}

\def\mf{{\rm I}}
\def\ms{{\rm I\hskip -0.5mm I}}

\def\ri{\rightarrow}

\def\pz{\partial_z}
\def\pzb{\partial_{\bar z}}
\def\py{\partial_y}
\def\pyb{\partial_{\bar y}}
\def\cpyb{\nabla_{\bar y}}
\def\cpybf{\nabla_{\bar y_1}}
\def\cpybs{\nabla_{\bar y_2}}

\def\dyb{\delta (\bar y -\bar y')}

\def\omf{\tilde \omega_\mf}
\def\oms{\tilde \omega_\ms}

\def\mtt{\tilde t}
\def\mto{\tilde \omega}

\hoffset=1in
\hsize=6.5in
\voffset=.55in
\vsize=8.9in

\advance \abovedisplayskip -10pt
\advance \belowdisplayskip -10pt
\advance \abovedisplayshortskip -10pt
\advance \abovedisplayshortskip -10pt

\titlepage
\singlespace
\vskip 2cm
\titlestyle{\bf Conserved Monodromy
$R^T \mht\circ \mht\circ = \mht\circ \mht\circ R^T $
Algebra}
\titlestyle{\bf
 in the Quantum Self-Dual Yang-Mills System
}

\bigskip
\bigskip
\bigskip
\bigskip

\author{Ling-Lie Chau and Itaru Yamanaka}

{\it \noindent Department of Physics, University of California,
Davis, California~~95616\nextline}

\vfill

\abstract
We
find a conserved
monodromy matrix differential
operator $\mht \circ$  in the
quantum Self-Dual Yang-Mills
(SDYM) system
and
show that it satisfies the
 exchange algebra $R^T \mht\circ \mht\circ = \mht\circ
\mht\circ R^T $.
 From its two infinitesimal forms, we obtain the
 infinite conserved quantum
nonlocal-charge algebras and the  infinite conserved
    Yangian algebras.
It is remarkable that such conserved algebras exist in a
four-dimensional nontrivial  quantum field theory
with interactions.

\vfill
\endpage

\Ref\F {L. D. Faddeev, Les Houches, Session XXXIX, 1982, p.565, eds. J.-B.
Zuber
and R. Stora.
}
\Ref\CY{ L.-L. Chau and I. Yamanaka, Phys. Rev. Lett. \undertext{70} (1993)
1916; Phys.
Rev. Lett. \undertext{68} (1992) 1807.}
\Ref\wz{J. Wess and  B. Zumino, Phys. Lett. \undertext{37B} (1971) 95, Serge
Novikov,
 Uspekhi Matematicheskikh Nauk, \undertext{5} (1982) 3.}
\Ref\W {E.Witten, Comm. Math. Phys. \undertext{92} (1984) 455.}
\Ref\DV { H. J. De Vega, Phys. Rept. \undertext{103} (1984) 243.}
\Ref\D {V.G. Drinfel'd, in Proceedings of the International Congress of
Mathematicians, Berkeley, California (1987).}
\Ref\BL { D. Bernard and A. LeClair, Commun. Math. Phys. \undertext{142}
(1991) 99. }
\Ref\M{N.J. MacKay, Phys. Lett. \undertext{B281} (1992) 90. }
\Ref\GN{ D.J. Gross and A. Neveu, Phys. Rev. \undertext{D10} (1974) 3235.}

In two-dimensional (2-d) quantum solvable systems, revealing algebraic
structure of operators is
a key step to solving the system. Among the algebras, the exchange algebra
of quantum monodromy matrix $\mht$,
 plays an important role in the quantum inverse
scattering method. Monodromy matrix is the boundary value of the
 solution to the linear systems
 or the group-valued element constructed from the
conserved Lie-algebra-valued local currents which satisfy the Kac-Moody
affine algebra. It generates
conserved charges that lead to the complete solutions of some
 2-d systems.\attach{[1]}

 In Ref.[2], we had formulated the quantum
self-dual Yang-Mills
(SDYM) system
in terms of the
group-valued local field $\tilde J$. In this paper,
 we construct a conserved quantum monodromy matrix operator
$\tilde T \circ$
in this nontrivial four dimensional quantum field theory with interactions.
The monodromy matrix
 $\tilde T \circ$, besides being a quantum field operator,
contains differential operators. It
obeys a quantum exchange
 algebra $R^T \mht\circ \mht\circ = \mht\circ \mht\circ R^T $. (We use a
tilde
 on top of a letter to indicate that it is a
quantum field operator; and a circle after a letter to denote that it
contains
the differential operator.)
 Taylor-expanding $\mht \circ$ in its spectral
parameter, we obtain the conserved infinite quantum nonlocal charges and
their algebras;
Taylor-expanding first in its exponent in the spectral parameter
we obtain the conserved infinite Yangian charges and
their Yangian algebras.
These algebras in the  four-dimensional (4-d) SDYM quantum field theory have
many more
generalized features than those in two dimensions.

It is remarkable that such conserved algebras exist in a
4-d nontrivial  quantum field theory
with interactions.
Because of their generalized features, the techniques developed
in two dimensions to extract physical information are not directly
applicable.
It is important and interesting to eventually find out
the physical implications of these algebras.

{\it The Quantum SDYM System}:

First, we briefly review the quantum Self-Dual Yang-Mills theory.
As formulated in Ref.[2], it is characterized by a
quantum field
Hamiltonian
$$ \eqalignno{
\tilde H_{int} = &- \alpha \int\int\int d{\bar y} dz d{\bar z}
\{ Tr \{ (\pzb \tilde J^{-1})(\pz \tilde J)\}\cr
&+ \int^1_0 d\rho Tr\{ (\tilde J^{-1} \partial_\rho \tilde J )
[(\pzb \tilde J^{-1})(\pz \tilde J)-(\pz \tilde J^{-1})(\pzb \tilde J)]\}\},
&(1)}$$
where $\tilde J = \tilde J (y,\bar y, z,\bar z)$
is a $2 \times 2$ matrix operator field
depending on the 4-d coordinates $ y, \bar y, z, \bar z; $ and $ y $ is the
 time.
(In this theory, the $z$ and $\bar z$ are discretized. To save writing, we
use integration in $z$ and $\bar z$ to denote summation in these
variables.\attach{[2]})

A group-valued local quantum field $\tilde J$ satisfies the
fixed-time-$y$ exchange algebra
$$\tilde J_\mf \tilde J_\ms = 1_{\mf , \ms}\tilde J_\ms
 \tilde J_\mf R^J_{\mf,\ms}, \eqno(2)$$
where
$$ R^J_{\mf ,\ms} \equiv P_{\mf ,\ms} \{ q^{\Delta_t \epsilon(\bar y-\bar y')
\delta_{zz'}\delta_{\bar z \bar z'}}  t^q_{\mf, \ms}
-q^{\Delta_s \epsilon(\bar y-\bar y')
\delta_{zz'}\delta_{\bar z \bar z'}}
s^q_{\mf, \ms}\} ,  \eqno(3)$$
and
$$
\eqalign{
 t^q_{\mf, \ms} &= diag \{1,{1 \over q+q^{-1}}\pmatrix{ q & 1 \cr
                                                    1 &q^{-1}\cr} , 1\},
{}~~s^q_{\mf, \ms}=1_{\mf, \ms}-t^q_{\mf, \ms}, \cr
q &\equiv e^{ -(i\hbar / 2\alpha a^2)\delta_{zz'}\delta_{\bar z \bar z'}} ;
{}~~ P_{\mf ,\ms}=diag \{1, \pmatrix{ 0 & 1 \cr
                                 1 & 0 \cr} , 1\}.
}
\eqno(4)$$
Note that $P_{\mf ,\ms}$ is the exchange matrix for tensor space $\mf$ and
$\ms$,
 i.e. $P_{\mf ,\ms} A_\mf B_\ms = A_\ms B_\mf P_{\mf ,\ms}$.
The constant $ q \neq 1  $ is from the quantum $ \hbar \neq 0 $ effect. It
also depends  in a nontrivial way
upon the cutoff $a^2$ and the type of Hamiltonian characterized by
the parameter $\alpha$ .
The field
$\tilde J^{-1}$, which
satisfies the
 relation $\tilde J^{-1} \tilde J = 1 =
\tilde J \tilde J^{-1}$,
obeys the fixed-time-$y$ exchange algebra
$$\tilde J^{-1}_\ms  \tilde J_\mf  =
 \tilde J_\mf R^J_{\mf,\ms} \tilde J^{-1}_\ms.
 \eqno(5) $$
 The equation of motion  obtained from
$\py (  \tilde J \pyb \tilde J^{-1} ) =
i/\hbar [ \tilde H_{int},  \tilde J \pyb \tilde J^{-1} ] $
is
$$ \py (\tilde J \pyb \tilde J^{-1}) =
- \pz ( \tilde J \pzb \tilde J^{-1}).\eqno(6)$$

 From the above exchange algebra, Eq.(2),
 we showed in Ref.[2] that the current
$ \tilde  j \equiv \tilde J \pyb \tilde J^{-1}  $ satisfies a 4-d
fixed-time-y
Kac-Moody affine algebra with a central charge
$\delta'(\bar y_1 - \bar y_2)$
-term
$$ [ {\tilde j}_{\mf} , ~{\tilde j}_{\ms}
] = [ P_{\mf,\ms}, {\tilde j}_{\ms}  ] \delta(\bar y_1 -\bar y_2)
\delta_{z_1, z_2}\delta_{\bar z_1, \bar z_2}
+ P_{\mf,\ms} \delta'(\bar y_1 -\bar y_2)\delta_{z_1, z_2}\delta_{\bar z_1,
 \bar z_2} .
\eqno(7)$$
It is a relation true only at fixed time-y since
$\tilde j(y,z,\bar y,\bar z)$ is not conserved.

On the other hand, we can show\attach{[2]} that the z-integrated (summed)
current

\noindent
$\tilde {\cal J} (\bar y,\bar z)=\int^l_{-l}   \tilde j(z,\bar y,\bar z)~dz$
is conserved
by checking
 $\partial_y \tilde {\cal J} =1/\hbar [H_{int},
 \tilde J \pyb \tilde J^{-1} ]
=0$.
Summing Eq.(7) in $z_1$ and $z_2$, we obtain the
conserved
Kac-Moody affine algebra,
$$ [ ({\tilde \cpybf })_{\mf}\circ , ({\tilde \cpybs})_{\ms}\circ ] =
 [ P_{\mf\ms} \delta(\bar y_1 - \bar y_2),
{}~({\tilde \cpybs})_{\ms}\circ ] \delta_{\bar z_1, \bar z_2}, \eqno(8)$$
where
$$ \tilde \cpyb \circ \equiv 2l \pyb \circ + \tilde {\cal J}. \eqno(9)$$
Here the circle of $\pyb \circ$ signifies its differential-operator property,
$\partial_y \circ f(y)= \partial_y f(y) + f(y) \partial_y \circ$ or
$[\partial_y \circ , f(y)]=\partial_y f(y)$.
Notice that this affine algebra of $\tilde \cpyb \circ $
 has no central-charge $\delta'(\bar y_1 -\bar y_2)$-term.
This is crucial for deriving the exchange algebra
 of the
monodromy matrix operator
$ \mht \circ $,
$ R^T \mht \circ \mht \circ = \mht \circ \mht \circ R^T $,
 which we shall derive in the next section.

{\it The $R^T \mht\circ \mht\circ = \mht\circ \mht\circ R^T $
Exchange Algebra}:

Since there is no
central charge
term in
 Eq.(8), the $\tilde \cpyb \circ $'s at different points of $ \bar z$
commute and
 we can always define a path-ordered exponential function of
$ \tilde \cpyb \circ $
$$ \tilde \psi(\bar y, \bar z; \lambda) \circ \equiv \overleftarrow P
\exp ( \lambda \int^{\bar z}_{-\bar l} d\bar z \tilde \cpyb\circ ),
\eqno(10)$$
where $\lambda $ is an arbitrary complex number.
For $\bar z$ equal to the boundary value, it becomes the
monodromy matrix operator
$$ \tilde T \circ
=[\tilde \psi(\bar y, \bar z; \lambda)\circ ]_{\bar z=\bar l}.
 \eqno(11)$$
The path-ordered integration is the path-ordered product of the following
infinitesimal elements:
 $$
\eqalign{
\tilde L_n \circ &\equiv ( 1+\lambda \Delta {\bar z} \tilde \cpyb\circ ),
{}~~
\hbox{and} ~~
{\bar z}=n\Delta {\bar z}; \cr
 \mht \circ &= ( 1+\lambda \Delta {\bar z} \tilde \cpyb\circ )_{\bar z =\bar
l}
( 1+\lambda \Delta {\bar z} \tilde \cpyb\circ )_{\bar z =\bar l -
\Delta {\bar z}}
\cdots
( 1+\lambda \Delta {\bar z} \tilde \cpyb\circ )_{\bar z =-\bar l}  \cr
&  = \tilde L_{\bar z =\bar l} \circ
\tilde L_{\bar z =\bar l - \Delta {\bar z}} \circ
\cdots
\tilde L_{\bar z =-\bar l} \circ .} \eqno(12) $$
Notice we treat $\bar y$ and $\bar y'$ as independent variables, i.e., $\pyb
f(\bar y') =0$.
Using the
Kac-Moody
affine algebra of $\tilde \cpyb\circ$, Eq.(8), we can straightforwardly
show that the $\tilde L_n \circ $'s satisfy the following exchange algebras
 $$       R^T_{\mf , \ms}(\lambda , \mu) ~
[\tilde L_n(\lambda)]_\mf\circ ~
[\tilde L_n(\mu)]_\ms\circ
=[\tilde L_n( \mu)]_\ms\circ ~
[\tilde L_n(\lambda)]_\mf\circ ~ R^T_{\mf , \ms}(\lambda , \mu),
{}~~~for ~n =m, \eqno(13)$$
where
$$   R^T_{\mf , \ms}
(\lambda, \mu)
\equiv 1-P_{\mf , \ms}{\lambda\mu \over \lambda-\mu}
\delta(\bar y - \bar y') ; \eqno(14)$$
and
$$       (\tilde L_n)_\mf\circ (\tilde L_m)_\ms\circ
= (\tilde L_m)_\ms\circ (\tilde L_n)_\mf\circ ,  ~~~
for ~n \neq m. \eqno(15)$$
Using  the fact that $\tilde T \circ $ is the product of
$\tilde L_n \circ $, Eq.(12), we
then derive
the exchange algebra of the monodromy matrix,
$$   R^T_{\mf,\ms}(\lambda , \mu)~\tilde T_{\mf}(\lambda)\circ~
\tilde T_{\ms}(\mu)\circ
  = \tilde T_{\ms}(\mu)\circ~
\tilde T_{\mf}(\lambda)\circ~R^T_{\mf,\ms}(\lambda , \mu). ~~
   \eqno(16)$$
 Notice that it would not be possible to derive Eq.(16) from Eqs.(13)-(15),
if
there were
 a central-charge $\delta'(\bar y_1 -\bar y_2)$-term
 in the  Kac-Moody affine algebra of $\tilde \cpyb \circ$, Eq.(8).

Eq.(16) is the central result of this paper.  This 4-d exchange algebra
 has two generalizations when compared to those in two
dimensions\attach{[1]}:
the monodromy matrix contains differential operator
$\pyb \circ$ and it depends on coordinate
$\bar y$.

It is known that the conserved currents in the
quantum WZNW theory\attach{[3],[4]} and the non-conserved local currents in
the quantum principal
chiral model \attach{[5]} in two dimensions satisfy the affine algebra with
central charge
and no consistent exchange algebra for their monodromy matrix
can be constructed.
Therefore, it is interesting and remarkable that a conserved quantum algebra
 $ R^T \mht \circ \mht \circ = \mht \circ \mht \circ R^T $ exists
for a nontrivial 4-d quantum field theory with interactions.

{\it Infinite Conserved Nonlocal-Charge Algebras and Infinite Conserved
Yangian
Algebras}:

Let's first discuss the infinite conserved
non-local charge algebras.
We Taylor-expand $\tilde T(\lambda)\circ$ in $\lambda$
 and $\pyb\circ$
$$ \tilde T\circ = \sum_{n=0}^{\infty}
\sum_{m=n}^{\infty}  \tilde  t^{(m,n)}(\bar y)
 \lambda^m (\partial\circ )^n,  \eqno(17)$$
which defines the quantum nonlocal charges
$\tilde t^{(m,n)}(\bar y) . $
They have the following expressions
in terms of the field $\tilde J$:
$$\eqalignno{ \tilde t^{(0,0)} &=1, &(18)\cr
 \tilde t^{(1,0)} &= \int^{\bar l}_{-\bar l} d\bar z \tilde \cpyb \circ  1
=\int^{\bar l}_{-\bar l} d\bar z
\int^{l}_{- l} dz \tilde J \pyb \tilde J^{-1},  &(19)\cr
 \tilde t^{(2,0)} &=\int^{\bar l}_{-\bar l} d\bar z_2  \tilde \cpyb
\circ  \int^{\bar z}_{-\bar l} d\bar z_1
\tilde \cpyb \circ 1  ,
 &(20)\cr
\hbox{and} ~~~~~~~~~~ & \cr
 \tilde t^{(n,0)} &=[\int^{\bar l}_{-\bar l} d\bar z_n  \tilde \cpyb \circ ]
[\int^{\bar z_n}_{-\bar l} d\bar z_{n-1}  \tilde \cpyb \circ  ]
\dots [\int^{\bar z_2}_{-\bar l} d\bar z_1  \tilde \cpyb  ] \circ 1.
&(21)  }$$
Substituting Eq.(17) into Eq.(16),
we show that
the nonlocal-charges satisfy the following algebras:
$$ \eqalignno{[ \tilde t_\mf^{(1,0)} , \tilde t_\ms^{(1,0)} ]
&= [P_{\mf,\ms}, \tilde t_{\ms}^{(1,0)}]\dyb -
\tilde t_\ms^{(1,1)} \pyb \dyb P_{\mf,\ms},
  &(22)\cr
 [ \tilde t_{\mf}^{(2,0)} , \tilde t_{\ms}^{(1,0)} ]& =
[P_{\mf,\ms}, \tilde t_{\ms}^{(2,0)}]\dyb
- (\tilde t_{\ms}^{(2,1)} \pyb + \tilde t_{\ms}^{(2,2)} \pyb^2)\dyb
P_{\mf,\ms} , &(23) \cr
\hbox{and} ~~~~~~~~~~~~~ & \cr
 [\tilde t_{\mf}^{(n,0)}, \tilde t_{\ms}^{(m,0)}]
&= P_{\mf, \ms} (\Sigma_{i=0}^{\infty}
    \tilde t_{\mf}^{(n+i,0)} \tilde t_{\ms}^{(m-i-1,0)}) \dyb  \cr
    & ~~~  -P_{\mf, \ms} \Sigma_{p=0}^{\infty}
\Sigma_{q=0}^{\infty}
(\Sigma_{i=0}^{\infty} \tilde t_{\mf}^{(m-i-1,p)}
 \tilde t_{\ms}^{(n+i,q)})  \pyb^q \partial_{\bar y'}^p \dyb ,  &(24)
  }$$
with  positive integers  $m, n, p$.
In Eq.(24),
 we have defined $\tilde t^{(n,p)}=0$ for  $ n<p $.
The first nonlocal charge algebra, Eq.(22), is nothing but the
$z$- and $\bar z$-summed
Kac-Moody
affine algebra of
Eq.(7).

Next,
we  make the $\lambda$-expansion of the monodromy matrix first in the
exponent
and obtain the Yangian charges and their algebra:

$$ \tilde T\circ \equiv \sum^{\infty}_{n=0}
[ \exp{( \sum^{\infty}_{m=n}  \mto^{(m,n)} \lambda^m )} ] (\pyb\circ)^n,
\eqno(25)$$
where the $ \mto^{(m,n)}$'s are the Yangian charges and they satisfy
the following algebras
$$
\eqalignno{
[\omf^{(1,0)} , \oms^{(1,0)}] &=
( [ P_{\mf,\ms}, \omf^{(1,0)}] \dyb - \omf^{(1,1)} \pyb \dyb P_{\mf,\ms} )
,  &(26)\cr
 [ \omf^{(2,0)}, \oms^{(1,0)} ] &= [ P_{\mf,\ms} , \omf^{(2,0)} ] \dyb - \cr
 \{ ( \omf^{(2,1)} +{1\over 2} (\omf^{(1,1)})^2 &-{1\over 2} \omf^{(1,0)}
\omf^{(1,1)} -{1\over 2} \omf^{(1,1)}\oms^{(1,0)} ) \pyb + \omf^{(2,2)}
\pyb^2
\}
\dyb P_{\mf,\ms} , \cr
& &(27) } $$
and etc.

For these infinite algebras we have not
found an explicit general expression for the higher order terms,
as we have done for the previous case, Eq.(24).

These algebras
have many generalized new features compared to those
in
 two dimensions\attach{[6,7,8]}:  (1) the
charges  $\tilde t$'s and $\mto$'s depend on coordinates
$\bar y$; (2) there are the central charge $\pyb^{(n)} \dyb$-terms.

The  $\mto^{(m,n)}$'s and the $\mtt^{(m,n)}$'s  are related
$$ \eqalignno{
\mtt^{(1,n)} &=\mto^{(1,n)}, &(28) \cr
 \mtt^{(2,n)}&=\mto^{(2,n)}+{1\over 2} (\mto^{(1,n)})^2, ~~etc. &(29)\cr}$$

 {\it Co-product} :

Motivated by Drinfel'd's's work\attach{[6]}, we define co-product for our
algebras:

$$ \Delta (\tilde T_\mf(\lambda)\circ) \equiv
\tilde T^A_\mf (\lambda)\circ \tilde T^B_\mf (\lambda)\circ , \eqno(30)$$
where super $A,B$ stands for the different Fock spaces
the  operators $\tilde T \circ $ acts on; thus,
 in the right hand side (r.h.s.) of Eq.(30), the product of two
monodromy matrices is a product merely in the matrix sense (not in the
quantum-field-operator sense since the two operators act on different Fock
spaces
and they
 commute).

We shall show that we can introduce the following distributive rule for the
co-product
$$ \Delta(\tilde T_\mf\circ \tilde T_\ms\circ) = \Delta(\tilde T_\mf)\circ
 \Delta(\tilde T_\ms)\circ, \eqno(31)$$
and that it is consistent with the exchange relation, Eq.(16).
Taking the co-product of both sides of Eq.(16), we obtain
$$ R^T_{\mf, \ms} \Delta(\tilde T^A_\mf (\lambda))\circ
 \Delta (\tilde T^A_\ms (\mu))\circ
= \Delta (\tilde T^A_\ms (\mu))\circ
\Delta (\tilde T^A_\mf (\lambda))\circ R^T_{\mf, \ms};
\eqno(32)$$
on the other hand, we need to check that this equality is true directly from
 the definition of co-products:

\noindent
l.h.s. of Eq.(32)

\noindent
$ = R^T_{\mf, \ms} \Delta (\tilde T^A_\mf (\lambda))\circ
 \Delta (\tilde T^A_\ms (\mu))\circ$;
{}~~~ from the definition of co-products,

\noindent
$ = R^T_{\mf, \ms} \tilde T^A_\mf (\lambda)\circ \tilde T^B_\mf
(\lambda)\circ
\tilde T^A_\ms (\mu)\circ
\tilde T^B_\ms (\mu)\circ$;
 ~~~ from $\tilde T^B_\mf (\lambda)\circ \tilde T^A_\ms (\mu)\circ =
\tilde T^A_\ms (\mu)\circ \tilde T^B_\mf (\lambda)\circ$,

\noindent
$ = R^T_{\mf, \ms} \tilde T^A_\mf (\lambda)\circ
 \tilde T^A_\ms (\mu)\circ \tilde T^B_\mf (\lambda)\circ
\tilde T^B_\ms (\mu)\circ$; ~ using the exchange algebra, Eq.(6) twice,

\noindent
$ =\tilde T^A_\ms (\mu)\circ \tilde T^A_\mf (\lambda)\circ
 \tilde T^B_\ms (\mu)\circ \tilde T^B_\mf (\lambda)\circ
R^T_{\mf, \ms}$; ~~~ using $\tilde T^A_\mf (\lambda)\circ
 \tilde T^B_\ms (\mu)\circ
=\tilde T^B_\ms (\mu)\circ \tilde T^A_\mf (\lambda)\circ$,

\noindent
$=\tilde T^A_\ms (\mu)\circ
 \tilde T^B_\ms (\mu)\circ
 \tilde T^A_\mf (\lambda)\circ \tilde T^B_\mf (\lambda)\circ
R^T_{\mf, \ms}$; ~~~ from the definition of co-product,

\noindent
$= \Delta (\tilde T^A_\ms (\mu))\circ
  \Delta (\tilde T^A_\mf (\lambda))\circ R^T_{\mf, \ms}
=$ r.h.s. of Eq.(32) ; \hskip 4.7cm  (33)

\noindent
therefore, we now have demonstrated the consistency of our co-product.

\def\mtt{\tilde t}
Expanding Eq.(30), we obtain the co-products of the charges.

The co-products of the nonlocal charges are

$$ \eqalignno{
\Delta (\mtt^{(1,0)}) &= (\mtt^{(1,0)})^A + (\mtt^{(1,0)})^B , &(34)\cr
   \Delta (\mtt^{(2,0)}) &= (\mtt^{(2,0)})^A + (\mtt^{(2,0)})^B
+ (\mtt^{(1,0)})^A  (\mtt^{(1,0)})^B
+ (\mtt^{(1,1)})^A \pyb (\mtt^{(1,0)})^B , \cr
\hbox{and etc;}    & &(35)\cr
 & \cr
\Delta (\mtt^{(1,1)}) &= (\mtt^{(1,1)})^A + (\mtt^{(1,1)})^B ,  &(36)\cr
   \Delta (\mtt^{(2,2)}) &= (\mtt^{(2,2)})^A + (\mtt^{(2,2)})^B
+ (\mtt^{(1,1)})^A  (\mtt^{(1,1)})^B , &(37)  \cr
\hbox{and etc;}  & \cr}$$
and etc.

We have checked that these co-products for the non-local charges,
 Eqs.(34)-(37), are consistent with
the non-local-charge
algebras, Eqs.(22)-(24), in the same manner as we checked that Eq.(32) is
consistent with
Eq.(16).

The co-products of the $\mto^{(m,n)}$'s are
$$
\eqalignno{
\Delta (\mto^{(1,0)}) &= (\mto^{(1,0)})^A + (\mto^{(1,0)})^B , &(38) \cr
 \Delta (\mto^{(2,0)}) &= (\mto^{(2,0)})^A + (\mto^{(2,0)})^B
+{1 \over 2} [ (\mto^{(1,0)})^A , (\mto^{(1,0)})^B ]
+ (\mto^{(1,1)})^A \pyb (\mto^{(1,0)})^B , \cr
\hbox{and etc;}     & &(39)  \cr
& \cr
\Delta (\mto^{(1,1)}) &= (\mto^{(1,1)})^A + (\mto^{(1,1)})^B ,  &(40) \cr
\Delta (\mto^{(2,2)}) &= (\mto^{(2,2)})^A + (\mto^{(2,2)})^B
+ (\mto^{(1,1)})^A  (\mto^{(1,1)})^B  ,  &(41) \cr
\hbox{and etc;}  & \cr }$$
and etc. (In the last two co-products, we use
the  co-product rule for the parameter $l$,
$ \Delta (l) = 2l. $ )

Similarly, we can also checked that these co-products
for the Yangian charges, Eqs.(38)-(41), are consistent with
the Yangian
algebras, Eqs.(26)-(27).

\vfill
\endpage
{\it  Automorphism}:

The exchange algebra of monodromy matrix, Eq. (16), has an automorphism,
since the $R$-matrix,
Eq.(14),
is in invariant
under the following transformation of the spectral parameters,

$$ {1 \over \lambda} \ri {1 \over \lambda} - \nu ~~~ and ~~~
 {1 \over \mu} \ri {1 \over \mu} - \nu, \eqno(42)$$
where $\nu$ is an arbitrary complex number as are $\lambda$ and $\mu$.
This implies that given Eq.(16), the following equation is also true
$$    R^T_{\mf\ms}(\lambda , \mu)\tilde T_{\mf}({\lambda \over 1-\nu
\lambda})
\circ
\tilde T_{\ms}({\mu \over 1- \nu \mu})
\circ
  = \tilde T_{\ms}({\mu \over 1-\nu \mu})
\circ
\tilde T_{\mf}({\lambda \over 1-\nu\lambda})
\circ
R^T_{\mf\ms}(\lambda , \mu), \eqno(43)$$
with the same $ R^T_{\mf\ms}(\lambda , \mu)$ matrix as in Eq.(16). Expanding
out in $\lambda$ and $\mu$, we obtain the
automorphism in $\mtt$'s and $\mto$'s
$$
\eqalign{
\mtt^{(0,n)} &\ri \mtt^{(0,n)}_{\nu} = \mtt^{(0,n)},  \cr
\mtt^{(1,n)} &\ri \mtt^{(1,n)}_{\nu} = \mtt^{(1,n)},  \cr
\mtt^{(2,n)} &\ri \mtt^{(2,n)}_{\nu} = \mtt^{(2,n)}+ \nu\mtt^{(1,n)}, \cr
\mtt^{(3,n)} &\ri \mtt^{(3,n)}_{\nu} = \mtt^{(3,n)}+ 2\nu\mtt^{(2,n)} + \nu^2
\mtt^{(1,n)},
\cr
\hbox{and etc;}    &} \eqno(44)$$
$$
\eqalign{
\mto^{(1,n)} &\ri \mto^{(1,n)}_{\nu} = \mto^{(1,n)} , \cr
\mto^{(2,n)} &\ri \mto^{(2,n)}_{\nu} = \mto^{(2,n)}+ \nu\mto^{(1,n)} , \cr
\mto^{(3,n)} &\ri \mto^{(3,n)}_{\nu} = \mto^{(3,n)}+ 2\nu\mto^{(2,n)} + \nu^2
\mto^{(1,n)} , \cr
\hbox{and etc.}& \cr}
\eqno(45)$$
The $ \mtt^{(m,n)}_{\nu}$'s and the $ \mto^{(m,n)}_{\nu}$'s satisfy the same
nonlocal-charge
 algebras, Eqs.(22)-(24), and the same Yangian algebras, Eqs.(26)-(27),
respectively.

Our  algebras also have the automorphism
in $\bar y$ and $\bar z$ coordinates because the $R^T$-matrix depends only on

the difference of two $\bar y$ coordinates and is independent of $\bar z$.

\vfill
\endpage
{\it Discussions and Conclusion}:

The Yangian algebras in two dimensions have provided the powerful means in
constructing S matrices for some integrable systems, e.g., the
the sine-Gordon model\attach{[7]} and  Gross-Neveu model\attach{[8],[9]}.
 We can not
directly transfer the techniques developed
in two dimensions to four dimensions
to extract physical implications, since the the algebras we
have obtained in
four dimensions
 have many generalized features.
What is surprising is that a nontrivial 4-d quantum field theory like the
 SDYM system can have
these beautiful conserved algebras. It is a challenge to us to eventually
decode
the physical implications of these conserved algebras.

\endpage

\refout

\end